\documentclass[aps, prd, reprint, showpacs, superscriptaddress, twocolumn, nofootinbib, amsmath]{revtex4-2}    

\usepackage{graphicx}% Include figure files
\usepackage{dcolumn}% Align table columns on decimal point
\usepackage{bm}% bold math
\usepackage{natbib}
\usepackage{amsmath}
\usepackage{siunitx}
\usepackage [autostyle, english = american]{csquotes}
\MakeOuterQuote{"}
\usepackage{lipsum}
\usepackage{babel}
\usepackage{appendix}
\usepackage{xcolor}
\usepackage{soul}

\newcommand{\neff}{\ensuremath{N_{\rm eff} }}

\begin{document}
\title{Effects of an Intermediate Mass Sterile Neutrino Population on the Early Universe}
\author{Hannah Rasmussen}
\email{hrasmussen@sandiego.edu}
\affiliation{Department of Physics and Biophysics, University of San Diego, San Diego, CA 92110}

\author{Alex McNichol}
\affiliation{Department of Physics and Biophysics, University of San Diego, San Diego, CA 92110}

\author{George M.\ Fuller}
\affiliation{Center for Astrophysics and Space Sciences, University of California, San Diego, La Jolla, CA 92093}

\author{Chad T.\ Kishimoto}
\affiliation{Department of Physics and Biophysics, University of San Diego, San Diego, CA 92110}
\affiliation{Center for Astrophysics and Space Sciences, University of California, San Diego, La Jolla, CA 92093}

\date{\today}% It is always \today, today,

\begin{abstract}

The hot and dense early Universe combined with the promise of high-precision cosmological observations provide an intriguing laboratory for Beyond Standard Model (BSM) physics. We simulate the early Universe to examine the effects of the decay of thermally populated sterile neutrino states into Standard Model products around the time of weak decoupling. These decays deposit a significant amount of entropy into the plasma as well as produce a population of high-energy out-of-equilibrium active neutrinos. As a result, we can constrain these models by their inferred value of \neff, the effective number of relativistic degrees of freedom. In this work, we explore a variety of models with \neff{} values consistent with CMB observations, but with vastly different active neutrino spectra which will challenge the standard cosmological model, affect lepton capture rates on free nucleons, and may significantly affect Big Bang Nucleosynthesis (BBN).

\end{abstract}

\maketitle

\section{\label{sec:level1}Introduction}

Upcoming astrophysical observations driven by Stage-4 CMB observatories and 30-m class telescopes look to make unprecedented, high-precision observations of the Universe, providing exciting opportunities to further probe both the standard cosmological model as well as the Standard Model \cite{CMBS4}.  Moreover, these observations can serve a powerful role in constraining possible Beyond Standard Model (BSM) physics models that are difficult -- and sometimes impossible -- to probe in the terrestrial laboratory.  In this work, we explore the implications of the decay of massive neutral fermions ({\it e.g.}, ``sterile neutrinos'') into Standard Model particles that will heat the photon-electron-positron-baryon plasma during the weak decoupling and Big Bang Nucleosynthesis (BBN) epoch as well as produce non-thermal high-energy neutrinos and antineutrinos of all flavors.

Current CMB-inferred values of \neff{} and the upper bound on $\sum m_{\nu}$ \cite{Planck2018} are consistent with the standard cosmological value of \neff{} of 3.046 \cite{Mangano2005} and neutrino oscillation experiments \cite{Katrin2018}. Current CMB-inferred values of deuterium and helium are also consistent with standard cosmological predictions \cite{Planck2018}, while primordial lithium abundances continue to provide tension between the standard BBN model prediction and observation. Stage-4 CMB observations will improve precision on these inferred values, and this improved precision will better probe the assumptions in the standard cosmological model while simultaneously constraining BSM physics and non-standard cosmological models. Improved precision on $\sum m_{\nu}$ from Stage-4 CMB observations may also result in upper bounds in tension with neutrino oscillation experiments \cite{CMBS4}. 30-m telescope observations will also improve the precision of observed primordial abundances and will further test the standard cosmological model.

Sterile neutrinos -- right-handed ``singlet'' neutrinos -- are a natural extension to the Standard Model to incorporate non-zero neutrino masses. For example, in the $\nu{\rm MSM}$, the introduction of sterile neutrinos can also address dark matter, baryon asymmetry, inflation, and electroweak symmetry while remaining consistent with neutrino oscillation experiments \cite{Davoudiasl2005,Asaka2005,st06}. Sterile neutrinos have been considered across several mass scales as dark matter candidates \cite{dw94,sf99a,pfkk16,kusenko06,Fuller2003,afp,dh02,Abazajian2002,Abazajian2006,abs05,st06,Boyanovsky2007a,Boyanovsky2007b,Shaposhnikov2007,Gorbunov2007,kf08,Laine2008,Petraki2008,pk07}, owing to the flexibility in their model and inherently non-interactive nature. They have also been proposed in mechanisms to provide the ``kicks'' that may explain high velocity pulsars \cite{Fuller2003,kusenko06,Kusenko1999,Barkovich2004,Loveridge2004,Kishimoto2011}, and in models that seek to explain the discrepancy between ``local Universe'' observations and CMB measurements of the Hubble constant \cite{Gelmini2021,Gelmini2020}. Sterile neutrinos have also been considered as enablers of the formation of the first stars \cite{Biermann2006,Mapelli2006,Stasielak2007}, as playing significant roles in core-collapse supernova explosions \cite{Hidaka2006,Fryer2006,hf07,Fuller2009,Suliga2019}, as taking part in baryogenesis \cite{Akhmedov1998,Asaka2005}, and as instigators of successful r-Process nucleosynthesis in neutrino-heated supernova ejecta \cite{McLaughlin1999,Caldwell2000,Fetter2003}. It has also been shown that sterile neutrino decay after the BBN epoch can initiate cascade nucleosynthesis \cite{Scherrer1984,Dolgov2000,Jedazmik2004,Jedazmik2006,Jedazmik2009,Kawasaki2005a,Cumberbatch2007,Pospelov2007,Ishiwata2010,Cyburt2010,Ellis2011,Pospelov2010,Pospelov2011}. In addition, results of several accelerator-based experiments have been interpreted as suggesting active-sterile flavor mixing \cite{Athanassopoulos1995,Athanassopoulos1996,Athanassopoulos1998a,Athanassopoulos1998b,Aguilar2001,Adamson2009,Aguilar2009,Aguilar2010,Abe2011,MiniBooNE2011a,MiniBooNE2011b}.

In this work, we study the effects of a sterile neutrino that can be thermally populated in the early Universe and then decays into Standard Model particles during the BBN epoch. Active-sterile neutrino mixing, parametrized by vacuum mixing angle $\theta \ll 1$, creates an effective weak interaction between the sterile neutrino and the rest of the Universe. However, this coupling is much weaker than the weak interaction, suppressed by a factor proportional to $\sin^2 \theta$. These sub-weak interactions can create the possibility of thermally populating the sterile states in the early Universe while also allowing for a long lifetime before their decay during the BBN epoch. This scenario was discussed in Ref.\ \cite{fkk11} and further refined in Ref.\ \cite{Gelmini2020}. To examine the impact of this model on BBN yields, we introduce a Boltzmann-like approach to evolve the neutrino and antineutrino distribution functions due to scattering and production in sterile neutrino decays. Throughout this paper we use the natural units, $c=\hbar=k=1$.

The number density of sterile neutrinos, $n_s$, can be written as 
\begin{equation}
 n_s = D\frac{3\zeta(3)}{2\pi^2}T_{\rm cm}^3e^{-t/\tau_s}.
 \label{eq:sterilenumberdensity}
\end{equation}
Here, $\zeta(3)$ is the Riemann Zeta function evaluated at 3, $\tau_s$ is the decay lifetime of the sterile neutrino, and $T_{\rm cm}$ is the comoving temperature, which scales inversely to the scale factor such that $T_{\rm cm}\propto a^{-1}$. The constant $D$ numerically describes the dilution of the sterile neutrinos from the time of their decoupling, prior to the QCD transition, to the time of weak decoupling described by our model. Since sterile neutrinos experience only sub-weak interactions with the plasma, they could be thermally populated and subsequently decouple from the plasma at temperatures much higher than weak decoupling, perhaps at the $\sim$ GeV scale \cite{fkk11}. The sterile neutrino number density is proportional to $T_{\rm cm}^3$ as would be expected from a decoupled, previously thermal species, as opposed to $T^3$ for a relativistic species in thermal equilibrium. $D$ captures the effects of dilution that occur as the effective degrees of freedom change, especially at the QCD transition, and can be inferred from the ratio of the comoving and plasma temperatures when the sterile neutrinos decouple and at weak decoupling:
\begin{equation}
    D = \left(\left. \frac{T}{T_{\rm cm}}\right\vert_{\rm \nu_s \text{-} {\rm dec}} \cdot \left. \frac{T_{\rm cm}}{T} \right\vert_{\rm wd} \right)^3 = \left(\frac{g_{\rm wd}}{g_{{\rm \nu_s \text{-} dec}}}\right) \approx \left(\frac{10.75}{61.75}\right).
    \label{eq:D}
\end{equation}
We take the degrees of freedom at the time of sterile neutrino decoupling, $g_{\rm \nu_s \text{-} dec}$, to include three quarks (up, down, and strange), gluons, pions, muons, electrons, neutrinos, and photons. The degrees of freedom at the time of weak decoupling, $g_{\rm wd}$, include only photons, electrons, positrons, and neutrinos.
%We take the degrees of freedom at the time of sterile neutrino decoupling, $g_{\rm \nu_s \text{-} dec}$, to include strange, up, and down quarks; gluons; pions; muons; electrons; neutrinos; and photons. The degrees of freedom at the time of weak decoupling, $g_{\rm wd}$, include only photons, electrons, positrons, and neutrinos.

The paper is organized according to the following. Section II will discuss the early Universe and how its dynamics are affected by the decaying sterile neutrinos. Section III will discuss the decay paths of the sterile neutrino and our Boltzmann-like approach to model the decay products and scattering. Section IV will present our results. Section V will provide a discussion of the implications of our model in the context of future terrestrial and astrophysical experiments as well as opportunities for future computational work.

\section{\label{sec:level2}The Early Universe}
The early Universe is comprised of a high-entropy plasma of photons, electrons, positrons, neutrinos, and baryons. However, baryons do not significantly contribute to the dynamics of the Universe at this epoch, as the baryon-to-photon ratio is on the order of $10^{-10}$. We simulate the expanding early Universe using the scale factor, $a$, as the independent variable with time, $t$, and plasma temperature, $T$, as dependent variables.

The expansion of the Universe is described by the Friedmann Equation,
\begin{equation}
  \left(\frac{1}{a}\frac{da}{dt}\right)^2 = \frac{8\pi}{3m_{\rm Pl}^2}\rho,
\end{equation}
where $m_{\rm Pl}$ is the Planck mass. The sum of the energy densities of all constituent particles of the early Universe, $\rho$, in our model is
\begin{equation}
 \rho = \rho_{\gamma}+\rho_{e^\pm}+\rho_{\nu}+\rho_{\nu_s}.
 \label{eq:energydensity}
\end{equation}
 We neglect to include baryons and dark matter because they both make negligible contributions to total energy density during this radiation-dominated epoch. 
 
 The Friedmann equation can be rearranged to show that
\begin{equation}
 \frac{dt}{da} = \sqrt{\frac{3m^2_{\rm Pl}}{8\pi}}\frac{\rho^{-1/2}}{a}.
 \label{eq:dtda}
\end{equation}
The energy density of photons is
\begin{equation}
 \rho_{\gamma} = \frac{\pi^2}{15}T^4.
 \label{eq:photonenergydensity}
\end{equation}
The energy density of positrons and electrons is
\begin{equation}
\begin{split}
 \rho_{e^\pm} &= \frac{2}{\pi^2}\int_0^\infty\frac{p^2\sqrt{p^2+m_e^2}}{e^{\sqrt{p^2+m_e^2}/T}+1}dp \\ &= \frac{2T^4}{\pi^2}\int_0^\infty\frac{\xi^2\sqrt{\xi^2+x^2}}{e^{\sqrt{\xi^2+x^2}}+1}d\xi,
 \label{eq:e_energydensity}
 \end{split}
\end{equation}
where the latter expression is made with the dimensionless variables $\xi=p/T$ and $x=m_e/T$. The energy density of the active neutrino population is 
\begin{equation}
 \rho_{\nu} = \frac{T_{\rm cm}^4}{2\pi^2}\int_0^{\infty} \epsilon^3f\left(\epsilon\right)d\epsilon.
 \label{eq:neutrinoenergydensity}
\end{equation}
The occupation fraction of neutrinos, $f(\epsilon)$, is a function of the scaled neutrino energy, 
\begin{equation}
    \epsilon = E/T_{\rm cm}.
    \label{eq:epsilon}
\end{equation} 
The scaled neutrino energy is introduced because the evolution of $f(\epsilon)$ is unaffected by the expansion of the Universe. The neutrino occupation fractions, $f(\epsilon)$, are dramatically affected by the products of $\nu_s$ decays and the concomitant scattering of these active neutrinos with the plasma, so a binned-spectrum of $f(\epsilon)$ must be included amongst our dependent variables. Finally, the energy density of the sterile neutrinos is simply
\begin{equation}
 \rho_{\nu_s} = m_sn_s.
 \label{eq:sterileenergydensity}
\end{equation}
 
The early Universe is homogeneous and isotropic, so only time-like heat flows can change the entropy in a comoving volume, $S \propto sa^3$ where $s$ is the entropy density of the electromagnetic plasma. Therefore,
\begin{equation}
    \frac{d}{da}(sa^3)=\frac{1}{T}\frac{dQ}{da},
    \label{eq:comovingentropy}
\end{equation}
 where $dQ/da$ is the rate at which thermal energy is deposited into a comoving volume of the plasma by out-of-equilibrium $\nu_s$ decays and scattering. As the sterile neutrinos decay, their rest mass energy is converted into energy in the active neutrino seas as well as in the plasma, so
\begin{equation}
    \frac{dQ}{da}=\frac{m_sn_sa^3}{\tau_s}\frac{dt}{da}-\frac{T_{\rm cm}^4a^3}{2\pi^2}\int \frac{df}{da}\epsilon^3 d\epsilon
    \label{eq:dQda}
\end{equation}
where $df/da$ is the derivative of the occupation fraction of neutrinos. This heating rate of the plasma is equal to the rate at which rest mass energy in sterile neutrinos is converted to Standard Model particles through decays minus the rate of energy going into the active neutrino seas. In the standard cosmological model, $dQ/da$ is slightly less than zero because as electrons and positrons annihilate into neutrino-antineutrino pairs and neutrinos undergo out-of-equilibrium up-scattering with the plasma, entropy is generated and energy is transferred from the plasma to the neutrino seas \cite{BURST}.

The total entropy density of the plasma is
\begin{equation}
    \begin{split}
    s &= s_\gamma + s_{e^\pm} \\
    & = \frac{4\pi^2}{45}T^3 + \frac{(\rho_{e^\pm}+P_{e^\pm})}{T}.
    \end{split}
    \label{eq:plasmaentropydensity}
\end{equation}
The pressure of the electrons and positrons is described by 
\begin{equation}
    \begin{split}
     P_{e^\pm} &= \frac{2}{3\pi^2}\int_0^\infty\frac{p^4}{\sqrt{p^2+m_e^2}(e^{\sqrt{p^2+m_e^2}/T}+1)}dp\\
     &= \frac{2T^4}{3\pi^2}\int_0^\infty\frac{\xi^4}{\sqrt{\xi^2+x^2}(e^{\sqrt{\xi^2+x^2}}+1)}d\xi,
    \label{eq:e_pressure}
    \end{split}
\end{equation}
and the energy density of the electrons and positrons is defined in Eq.\ (\ref{eq:e_energydensity}). Using these definitions, $dT/da$ is 
\begin{widetext}
\begin{equation}
    \frac{dT}{da} = \frac{\dfrac{1}{T}\dfrac{dQ}{da}- 3a^2\left[\dfrac{4\pi^2}{45}T^3 + \dfrac{1}{T}\bigg(\rho_{e^{\pm}}+P_{e^{\pm}}\bigg)\right]}{a^3 \left[ \dfrac{4\pi^2}{45}T^2 + \dfrac{1}{T} \left( \dfrac{d\rho_{e^\pm}}{dT} + \dfrac{dP_{e^\pm}}{dT} \right) - \dfrac{1}{T^2} \bigg( \rho_{e^{\pm}} + P_{e^{\pm}}\bigg) \right]}.
    \label{eq:dTda}
\end{equation}
\end{widetext}

\section{\label{sec:level3}Sterile Neutrino Decay}
In our model, we consider several different paths for the sterile neutrino to decay into into Standard Model products. The electromagnetic products thermalize in the plasma while the neutrinos and antineutrinos can exchange energy with the plasma through collisions. To computationally model how the active neutrino spectra evolves over time, we calculate $df/da$ for each decay process and scattering,
\begin{equation}
    \frac{df}{da} = \left. \frac{df}{da} \right\vert_{\nu_s\rm\,decay} + \left. \frac{df}{da} \right\vert_{\rm scattering}.
    \label{eq:dfda1}
\end{equation}

In the standard cosmological model, active neutrinos have a nearly thermal distribution after weak decoupling and electron-positron annihilation. However, when sterile decays introduce a population of highly energetic neutrinos to the neutrino seas, the neutrinos will no longer have a thermal distribution. Neutrino-neutrino collisions and neutrino-electron/positron collisions push the population towards thermal equilibrium, but the rate of these collisions is not rapid enough to rethermalize the population before weak decoupling is complete. The Appendix outlines $df/da$ from each individual decay path that we consider. In this section, we simply examine the decay rates of the decay processes included in our model and consider their general effects on the active neutrino population.

\subsection{\label{sec:level3a}Intermediate Sterile Neutrino Mass Decays}

\begin{figure} 
\includegraphics[scale=.28]{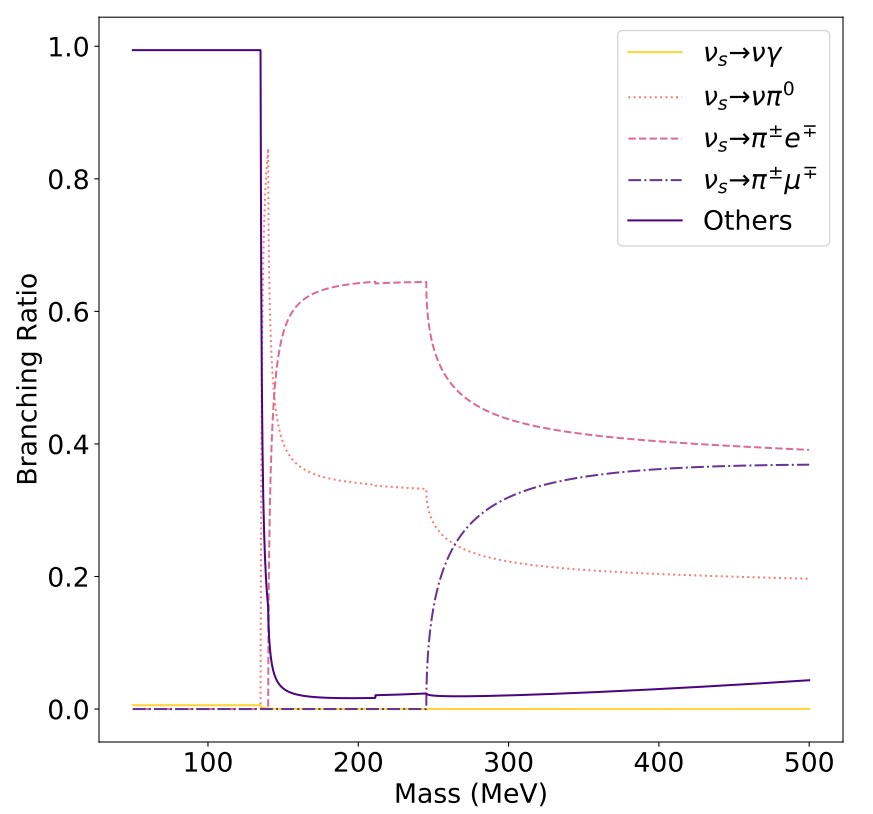}
\caption{Branching ratios of the four sterile neutrino decay paths included in our model as a function of mass. The solid line is the branching ratio of decay paths not included in our model. In this work we use $m_s = 300$ MeV, which is well approximated by decays 1-4.}
\label{fig:Branchingratio}
\end{figure}

We chose to include the following sterile neutrino decay processes in our calculations, as either their branching ratios are significantly higher than other decay processes available to the sterile neutrino in the mass regime of 150-500 MeV or their products have a noteworthy impact on the active neutrino population \cite{Abazajian2001,Banger1995}. The branching ratios of the decay processes we consider as a function of sterile neutrino mass can be visualized in Fig.\ \ref{fig:Branchingratio}. The first of four decay processes we consider in our model,
\begin{equation}
 1.\;\;\; \nu_s \rightarrow \nu + \gamma,
 \label{eq:decay1}
\end{equation}
is available to sterile neutrinos of all masses and has a branching ratio that remains below 0.1\% for our masses of interest. The sterile neutrinos that undergo this decay will produce monoenergetic high energy active neutrinos that will have a significant effect on the overall energy distribution of active neutrinos. This process has a decay rate of
\begin{equation}
 \Gamma_{1} = \frac{9\alpha G_F^2}{512\pi^4} m_s^5 \sin^2\theta,
 \label{eq:rate1}
\end{equation}
where $\alpha$ is the fine structure constant and $G_F$ is the Fermi constant \cite{Abazajian2001,Banger1995}. 

The second decay path we consider begins to account for a significant portion of sterile neutrino decays past the neutral pion mass of 135 MeV, and is thus important for all of our sterile neutrino models. The decay
\begin{equation}
 2.\;\;\; \nu_s \rightarrow \pi^0 + \nu
 \label{eq:decay2}
\end{equation}
also produces a monoenergetic high energy neutrino \cite{Dolgov2000}. The neutral pion decays quickly into two photons,
\begin{equation}
 \pi^0 \rightarrow 2\gamma,
 \label{eq:neutralpiondecay}
\end{equation}
and thus completely thermalizes, adding significant energy to the plasma. Decay 2 has the decay rate
\begin{equation}
 \Gamma_{2} = \frac{G_F^2f_\pi^2}{16\pi} m_s \left(m_s^2-m_{\pi^0}^2\right) \sin^2\theta
 \label{eq:rate2}
\end{equation}
where the pion decay constant $f_\pi = 131$ MeV \cite{Dolgov2000}.

Decay 3 follows the path \cite{Dib2018,Ballett2017}
\begin{equation}
 3.\;\;\; \nu_s \rightarrow \pi^\pm + e^\mp.
 \label{eq:decay3}
\end{equation}
Similarly, decay 3 accounts for a significant portion of decays where the sterile neutrino mass exceeds the sum of the charged pion mass and the electron mass of about 140 MeV. The charged pion decays further,
\begin{equation}
     \pi^\pm \rightarrow \mu^\pm + \nu_{\mu}.
 \label{eq:chargedpiondecay}
\end{equation}
Finally, the muon from the pion decay will decay quickly via the process
\begin{equation}
 \mu^\pm \rightarrow e^\pm + \nu_e + \bar{\nu}_\mu.
 \label{eq:muondecay}
\end{equation}
Thus, this decay produces a total of three active neutrinos. The decay rate of decay 3 is
\begin{equation}
   \begin{split}
        \Gamma_{3} =  \frac{2G_F^2f_\pi^2}{16\pi}m_s \biggl[\left(m_s^2-\left(m_{\pi^\pm}+m_e\right)^2\right)\\ \times \left(m_s^2-\left(m_{\pi^\pm}-m_e\right)^2\right)\biggr]^{1/2} \sin^2\theta.
   \end{split}
   \label{eq:rate3}
\end{equation}
where the factor of two accounts for the equal probability of $\nu_s \to \pi^+ + e^-$ and $\nu_s \to \pi^- + e^+$ \cite{Dib2018,Ballett2017}.

\begin{figure}
    \includegraphics[scale=0.265]{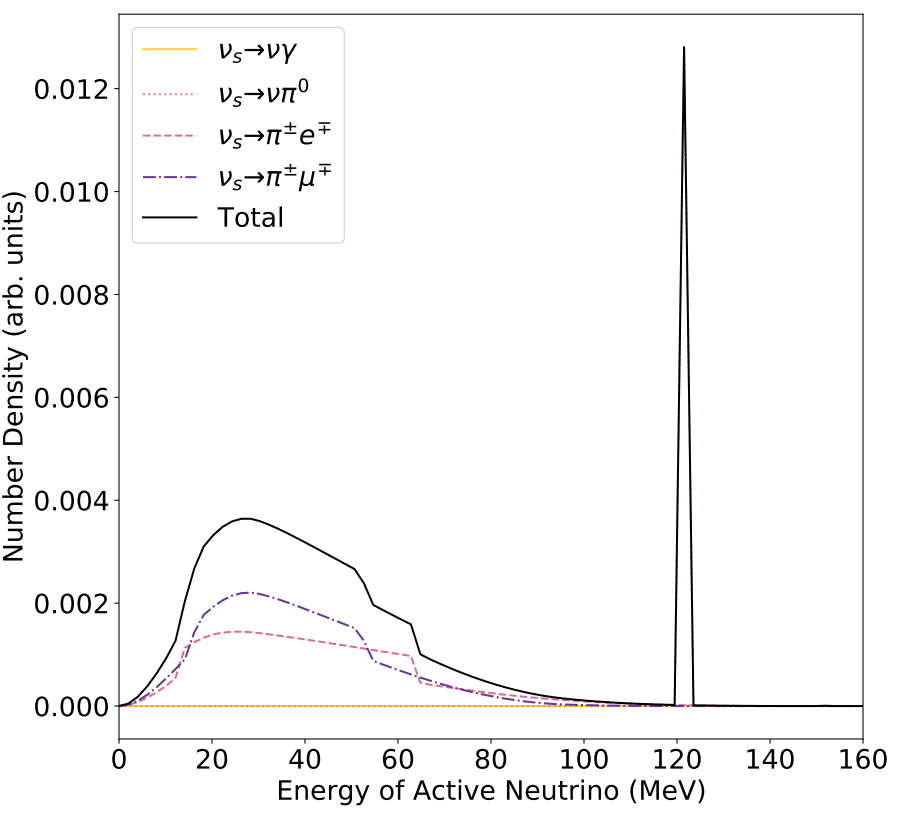}
    \caption{The active neutrino decay spectrum from sterile neutrino models with $m_s=300$ MeV and $\tau = 0.91$ s (henceforth the ``long-lived'' model, see Sec.\ \ref{sec:level4a}). The large peak in the total number density at approximately \mbox{$E=120$ MeV} is from decay 2. In addition, there is a peak from decay 1 at 150 MeV that is too small to see at this scale because the branching ratio for decay 1 is so low.}
    \label{fig:NumDensity}
\end{figure}

\begin{figure} 
\includegraphics[scale=.27]{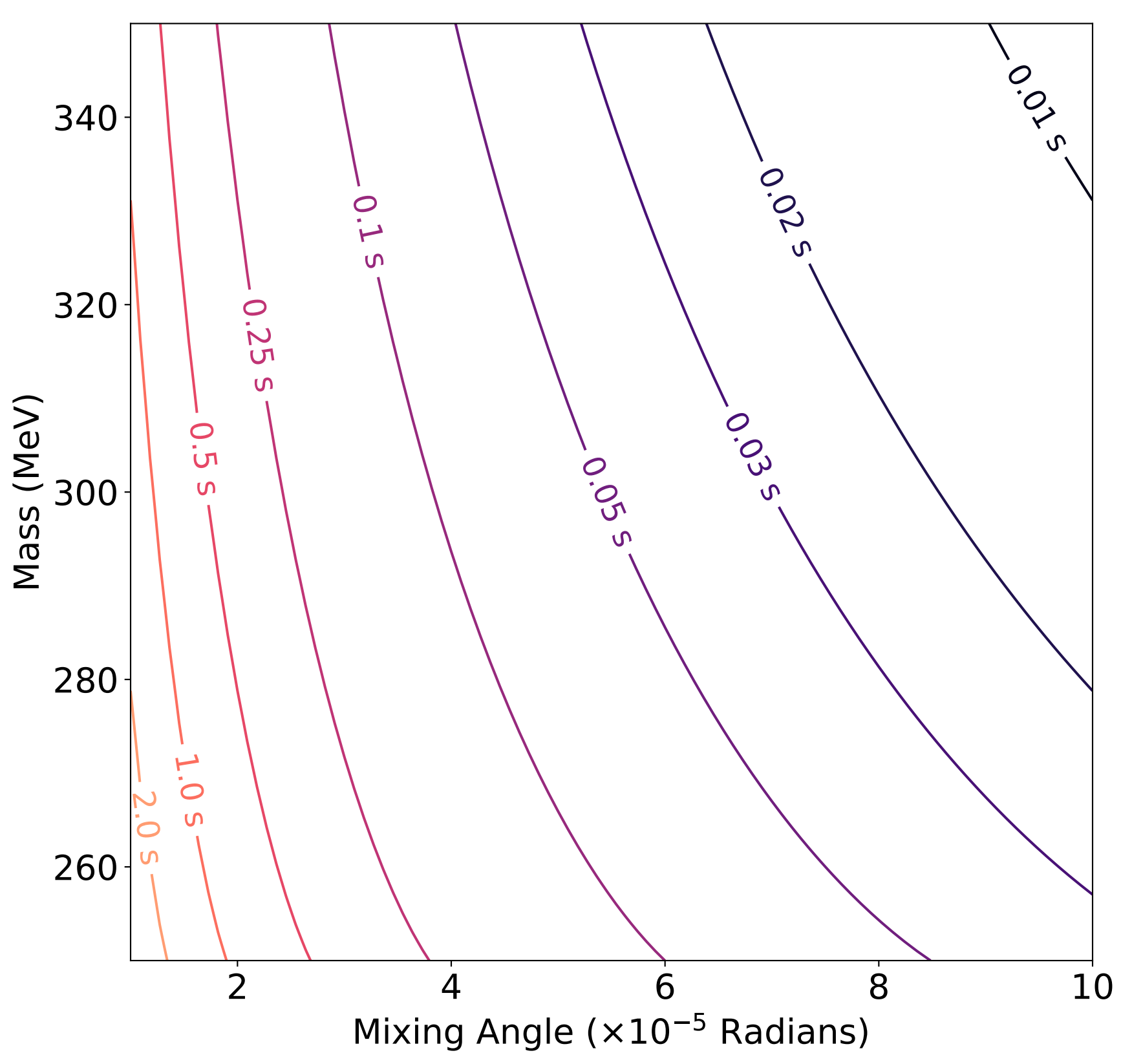}
\caption{Contours of sterile neutrino lifetimes as a function of mass and mixing angle.}
\label{fig:Lifetimes}
\end{figure}

Decay 4 instead creates muons and antimuons
\begin{equation}
 4.\;\;\; \nu_s \rightarrow \pi^\pm + \mu^\mp,
 \label{eq:decay4}
\end{equation}
but is only possible when $m_s \gtrsim 245$ MeV \cite{Dib2018,Ballett2017}. The decay rate is \cite{Dib2018,Ballett2017}
\begin{equation}
   \begin{split}
        \Gamma_{4} =  \frac{2G_F^2f_\pi^2}{16\pi}m_s \biggl[\left(m_s^2-\left(m_{\pi^\pm}+m_\mu\right)^2\right)\\ \times \left(m_s^2-\left(m_{\pi^\pm}-m_\mu\right)^2\right)\biggr]^{1/2} \sin^2\theta.
   \end{split}
   \label{eq:rate4}
\end{equation}
The charged pion follows the same decay path as the pion in decay 3, but in addition, the muon produced directly from the decay of the sterile will decay as well, producing an additional two neutrinos by Eq.\ (\ref{eq:muondecay}). Therefore, decay 4 produces a total of five active neutrinos. The spectra of neutrino products of decays 3 and 4 dominate the lower energy regime of neutrinos as shown in Fig.\ \ref{fig:NumDensity}. Henceforth, we parametrize the sterile neutrino models as a function of mass and lifetime, where lifetime of the sterile neutrino is inversely related to the sum of the individual rates
\begin{equation}
    \tau_s = \frac{1}{\Gamma_1+\Gamma_2+\Gamma_3+\Gamma_4}.
\end{equation}
The relation between mass, mixing angle, and lifetime is exhibited by Fig. \ref{fig:Lifetimes}.

\subsection{\label{sec:level3b}Other Sterile Neutrino Decays}

There are several other possible decay paths available to the sterile neutrino, but their contributions are either negligible in our mass range of interest or they don't have a significant effect on the neutrino distribution like decays 1 and 2. Decays that are more active at lower masses include decay processes into three leptons \cite{Abazajian2001,Banger1995,Ballett2017,Boyanovsky2014}. These processes include
\begin{equation}
 5.\;\;\; \nu_s \rightarrow 3\nu,
 \label{eq:decay5}
\end{equation}
\begin{equation}
 6.\;\;\; \nu_s \rightarrow \nu + e^+ + e^-,
 \label{eq:decay6}
\end{equation}
\begin{equation}
 7.\;\;\; \nu_s \rightarrow \nu + e^\pm + \mu^\mp,
 \label{eq:decay7}
\end{equation}
\begin{equation}
 8.\;\;\; \nu_s \rightarrow \nu + \mu^+ + \mu^-.
 \label{eq:decay8}
\end{equation}
We do not include these decays in our final calculations because within the range of $\nu_s$ masses of interest, the sum of the branching ratios of these decays remains negligible.

Higher mass sterile neutrinos experience decays into multiple hadrons, e.g.,
\begin{equation}
 9.\;\;\; \nu_s \rightarrow \pi^0 + \pi^\pm + l^\mp,
 \label{eq:decay9}
\end{equation}
\begin{equation}
 10.\;\;\; \nu_s \rightarrow 2\pi^0 + \nu,
 \label{eq:decay10}
\end{equation}
\begin{equation}
 11.\;\;\; \nu_s \rightarrow 2\pi^0 + \pi^\pm + l^\mp,
 \label{eq:decay11}
\end{equation}
\begin{equation}
 12.\;\;\; \nu_s \rightarrow 2\pi^\pm + \pi^\mp + l^\mp,
 \label{eq:decay12}
\end{equation}
\begin{equation}
 13.\;\;\; \nu_s \rightarrow 3\pi^0 + \nu,
 \label{eq:decay13}
\end{equation}
but they do not play a significant role in decays for $\nu_s$ masses less than 500 MeV \cite{Dib2018}.

\subsection{\label{sec:level3c}Collisions}
 Neutrino-neutrino and neutrino-electron/positron collision integrals are calculated using the results described in the Appendices of Ref.\ \cite{BURST}. The change in occupation fraction due to collisions is simply a sum of $df/da$ due to neutrino-neutrino collisions and $df/da$ due to neutrino-electron/positron collisions, so
 \begin{equation}
     \left. \frac{df}{da} \right\vert_{\rm scattering} = \left. \frac{df}{da} \right\vert_{\rm \nu \text{-} \nu \, scattering} + \left. \frac{df}{da} \right\vert_{\rm \nu \text{-} e^{\pm} \, scattering}
 \end{equation}
The neutrino-neutrino collision integrals are self-consistently calculated throughout. The neutrino-electron/positron collision integrals can be self-consistently calculated as well, but doing so makes the step-size in our code shrink by many orders of magnitude. This is due to the neutrino sea remaining largely in thermal equilibrium with the plasma for temperature regimes of interest due to the neutrino-electron/positron scattering in spite of unequal heating from the decaying sterile neutrinos. Such a computationally expensive calculation is beyond the scope of this work, so we introduce a simple analytic model for these collisions,
\begin{equation}
    \left. \frac{df}{da} \right\vert_{\rm \nu \text{-} e^{\pm} \, scattering} \approx -A n_e G_{\rm F}^2 p^n T^{2-n} \left(f-f_{\rm eq}\right),
\end{equation}
where $n_e$ is the number density of electrons and $f_{\rm eq}$ is the Fermi-Dirac distribution of the neutrinos if the neutrino sea were in thermal equilibrium with the plasma. The overall normalization, $A$, and the exponent, $n$, are fitting parameters in our model. Several times throughout our calculation, we first calculate the $\nu \text{-} e^{\pm}$ collision integral, then fit our model parameters $(A,n)$, and use this model to solve the ODEs. This provides a good balance between self-consistently calculating the collision integrals and solving the ODEs in a reasonable amount of time.

\section{\label{sec:level4}Results}

For our initial conditions, we set $T=T_{\rm cm}=15$ MeV, and we assume that the neutrino and antineutrino seas are in thermal equilibrium with the plasma and thus have a Fermi-Dirac thermal distribution, with temperature $T_{\rm cm}=T$, so
\begin{equation}
    f(\epsilon) = \frac{1}{e^\epsilon + 1}.
    \label{eq:occupationfraction}
\end{equation}
As described above, the scaled neutrino energy, $\epsilon = E/T_{\rm cm}$, is introduced because the evolution of $f(\epsilon)$ is unaffected by the expansion of the Universe. To solve the resulting ODEs, we use a Cash-Karp Runge-Kutta routine with an adaptive step size \cite{numrec}.

\begin{figure}
    \includegraphics[scale=0.34]{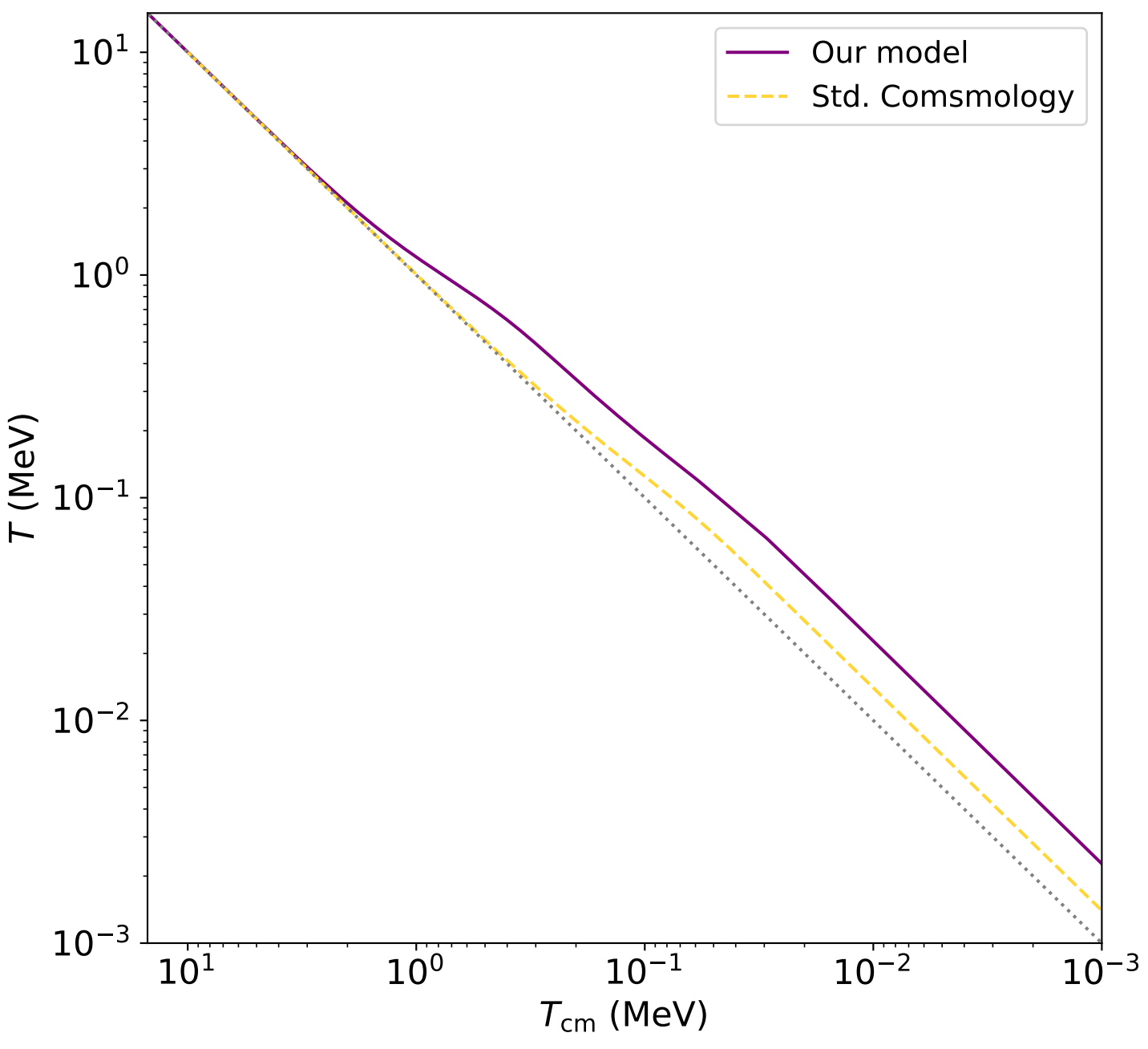}
    \caption{The evolution of plasma temperature in the long-lived model. For comparison, also shown is the plasma temperature in the standard cosmological model (dashed curve). A reference diagonal line (dotted line), $T=T_{\rm cm}$, is included to help identify when the plasma is heated by electron/positron annihilation in both models and when entropy is transferred to the plasma through the products of sterile neutrino decay.}
    \label{fig:TTcmLong}
\end{figure}

A decaying massive sterile neutrino population in the early Universe has several effects that arise in all of our models. One simple effect is that the plasma temperature is significantly boosted above what would be expected in a standard cosmological model due to the thermalization of the electromagnetic products of the sterile neutrino decays in the plasma and due to scattering between the high-energy active neutrino decay products and the plasma. This can be seen in Fig.\ \ref{fig:TTcmLong}. In the standard cosmological model, electrons and positons annihilate around $T_{\rm cm} \sim 0.1$ MeV, heating the plasma to the canonical result of $T=(11/4)^{1/3}T_{\rm cm}$. Fig.\ \ref{fig:TTcmLong} compares this canonical result to our model, in which the plasma is heated significantly more by the products of sterile neutrino decay. 

\begin{figure}
    \includegraphics[scale=0.35]{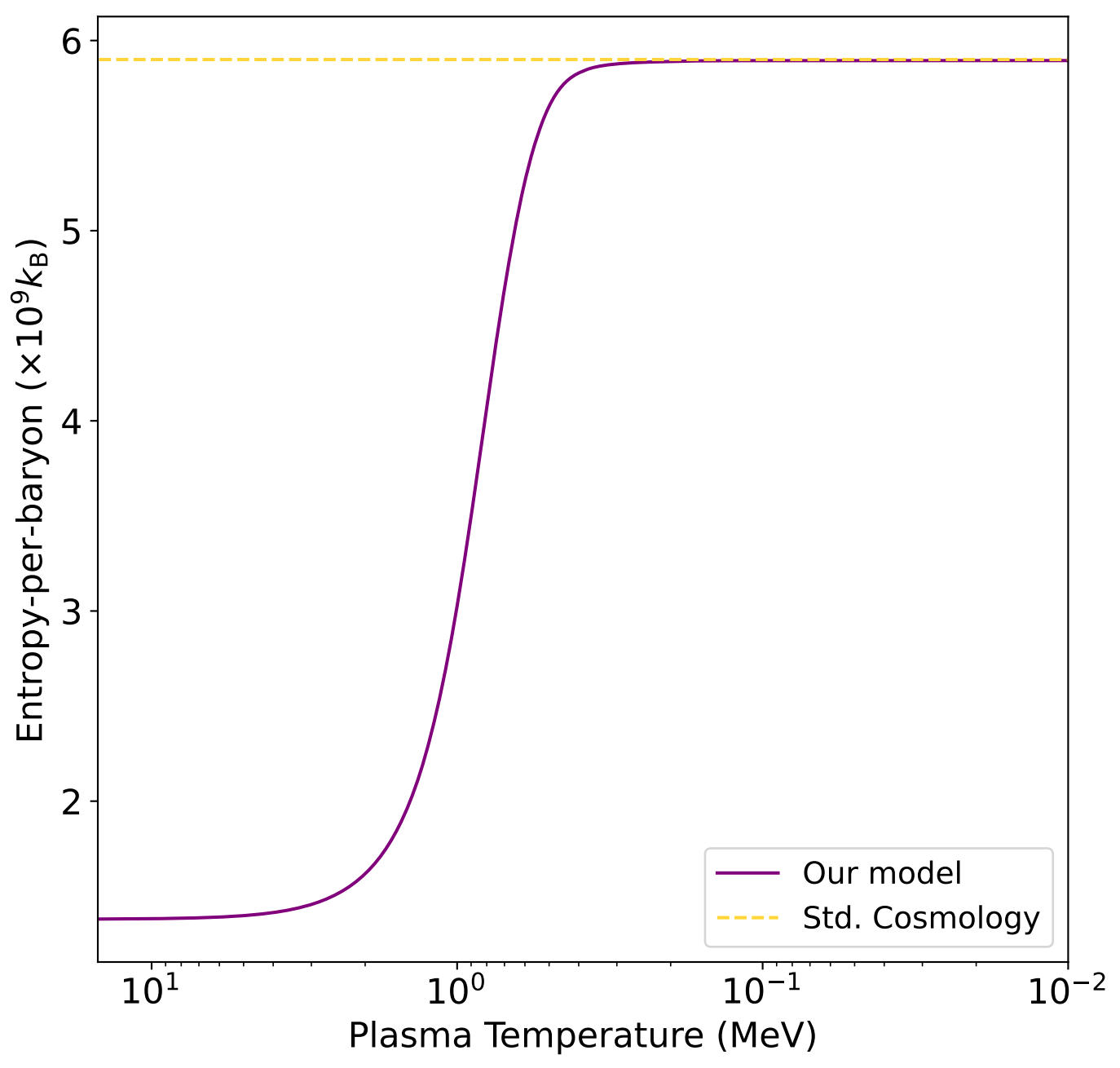}
    \caption{The entropy-per-baryon in the long-lived model. In standard cosmology, the entropy-per-baryon inferred from the CMB is approximately constant. In our model, the high energy decays of decoupled sterile neutrinos add entropy back into the plasma that cause an increase in the entropy-per-baryon.}
    \label{fig:EntropyPerBaryonLong}
\end{figure}

This heating of the plasma causes the dilution of species that are out of equilibrium with the plasma or are non-relativistic at the time of the decays. Fig.\ \ref{fig:EntropyPerBaryonLong} shows the evolution of the entropy-per-baryon in the plasma. In the standard cosmological model, the plasma is always nearly in equilibrium so the entropy-per-baryon remains constant at the value observed in the CMB. While it is true that in the standard cosmological model out of equilibrium scattering between neutrinos and the plasma {\it slightly} decreases the entropy of the plasma by a few tenths of a percent \cite{BURST}, sterile neutrino decays in our model add a significant amount of entropy to the plasma. So, matching the CMB-observed entropy-per-baryon at the time of recombination requires a much lower entropy-per-baryon at earlier epochs, interestingly during the weak decoupling and weak freeze-out epochs. 

\begin{figure}
    \includegraphics[scale=0.36]{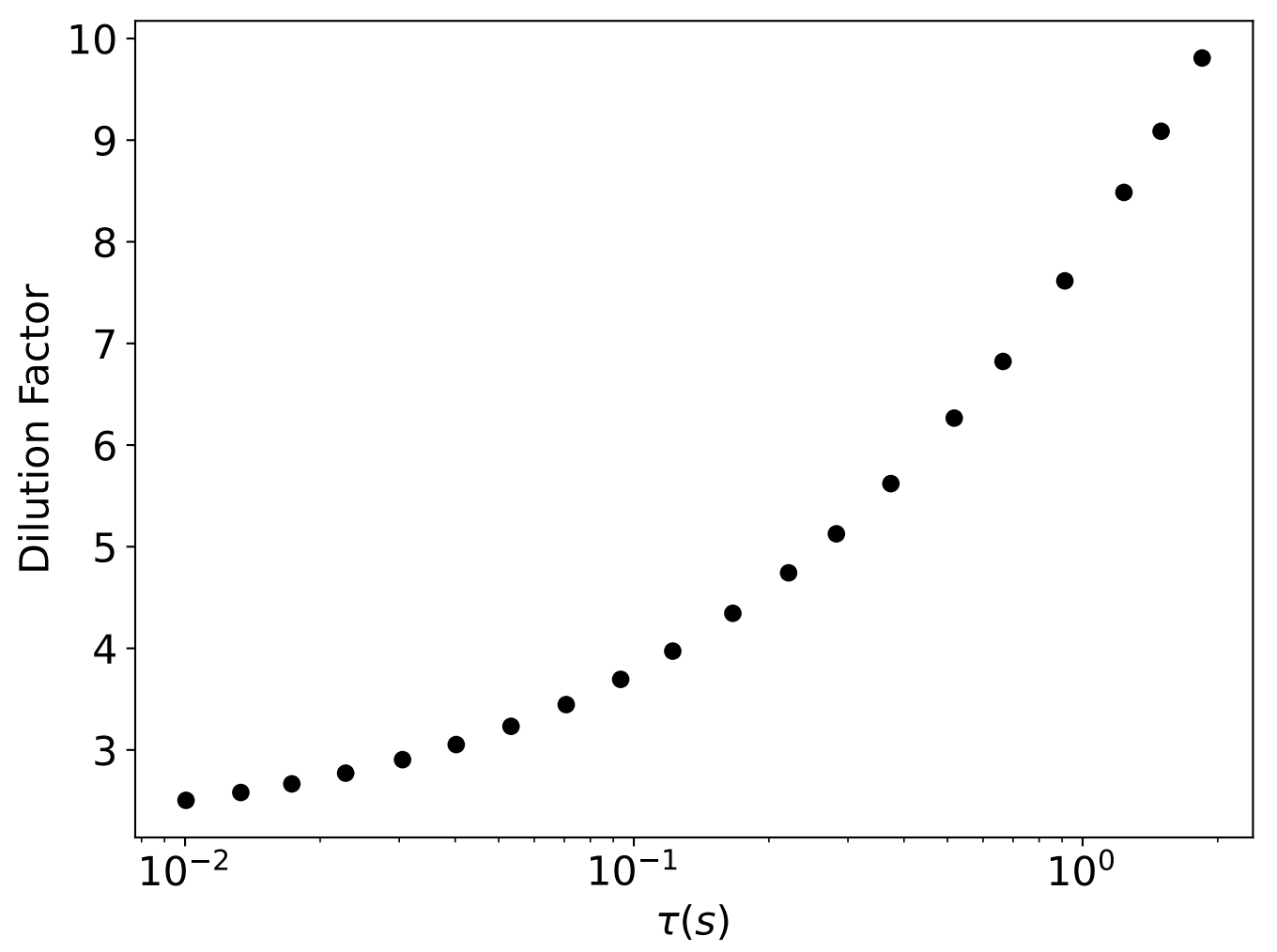}
    \caption{The dilution factors of sterile neutrinos with \mbox{$m_s = 300$ MeV} over varying lifetimes. As the sterile neutrinos decay later in time, their electromagnetic products are increasingly more energetic than the plasma at the time they decay, leading to the increase in dilution factor.}
    \label{fig:DilutionFactor}
\end{figure}

Sterile neutrino decays alter the active neutrino spectra in two principal ways: dilution and the production of high energy neutrinos. In the models of interest in this work, the active neutrinos begin to undergo weak decoupling around the time of these decays, so we expect the active neutrino distribution to experience some degree of dilution. Fig.\ \ref{fig:DilutionFactor} shows dilution factor as a function of sterile neutrino lifetime, where dilution factor is the ratio of final entropy-per-baryon to initial entropy-per-baryon,
\begin{equation}
    F \equiv \frac{S_{\rm final}}{S_{\rm initial}} = \frac{g_{\rm sf} T_{\rm f}^3 a_{\rm f}^3}{g_{\rm si} T_{\rm i}^3 a_{\rm i}^3}.
    \label{eq:dilutionfactor}
\end{equation}

\begin{figure}
    \includegraphics[scale=0.28]{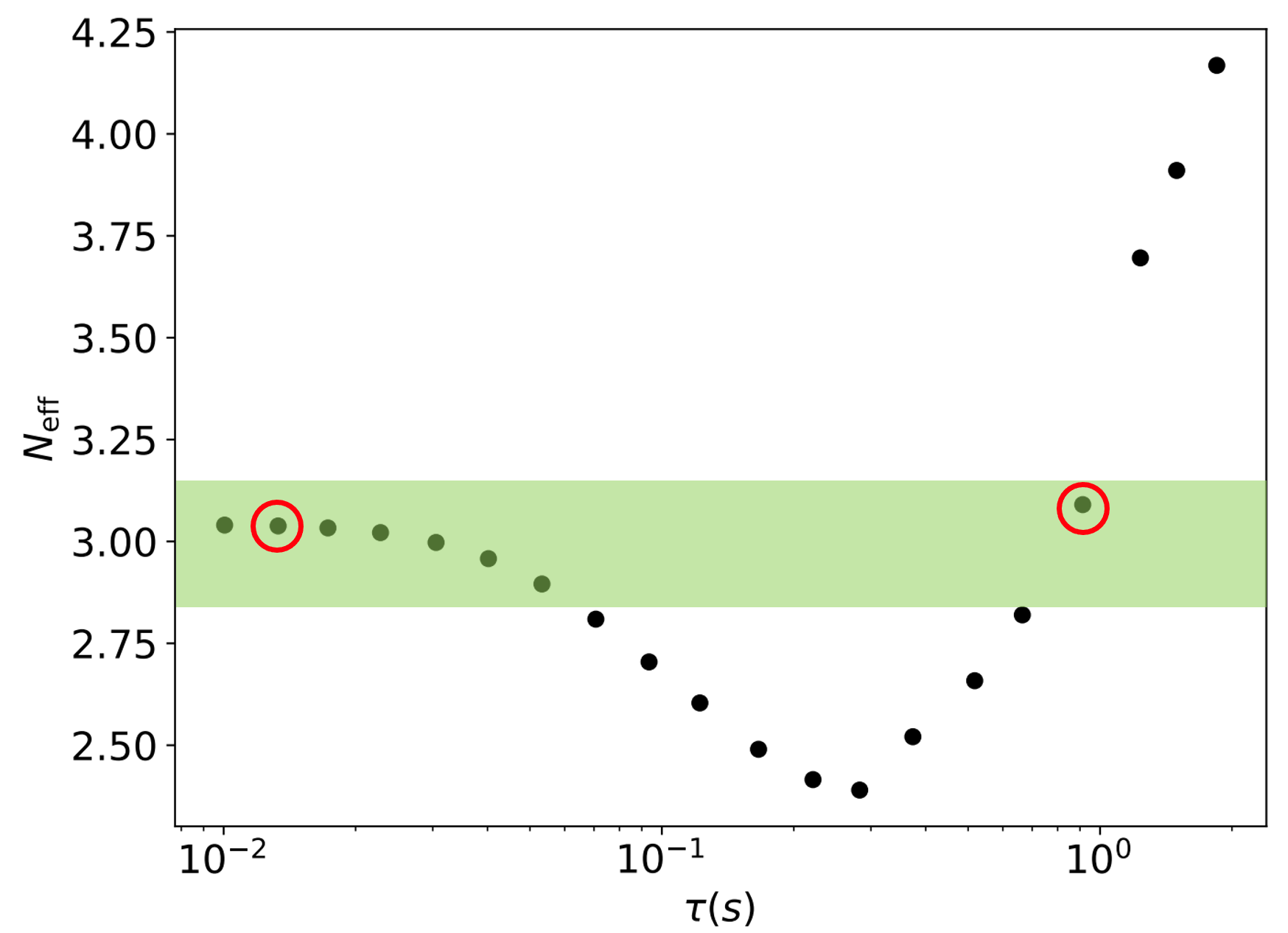}
    \caption{The \neff{} values of 300 MeV sterile neutrinos over varying lifetimes. The shaded green zone represents the most recent best-fit value of \neff{} from PLANCK, \neff $=2.99\pm0.17$ \cite{Planck2018}. A short-lived model and a long-lived model, circled in red, are both in the accepted range for \neff.}
    \label{fig:Neff}
\end{figure}

As lifetime of the sterile neutrino increases, the dilution factor increases due to an increase in the disparity between the plasma temperature and the energy of the electromagnetic products at the time they are dumped into the plasma from decay processes. Additionally, the population of high-energy neutrinos and antineutrinos will affect lepton capture rates, especially as they significantly overcome the energy threshold of $p + \bar{\nu}_e \to e^+ + n$. Alteration of these rates will affect the neutron-to-proton ratio. While performing a BBN network calculation is beyond the scope of this work, we note that BBN yields will be affected by the modifications to the entropy-per-baryon ratio, neutron-to-proton ratio, and the time-temperature relation. We leave a robust investigation of these effects to future work. We tested a range of sterile neutrino models with a mass of 300 MeV that decay with varying lifetimes roughly in the range coincident with the weak decoupling and weak freeze-out epochs in the early Universe. We use the final neutrino distribution, at $T \sim$ keV, to calculate \neff{} from the neutrino energy density, \begin{equation}
    \rho_{\nu} = \frac{7}{4}\left( \frac{4}{11}\right)^{4/3} \neff \frac{\pi^2}{30}T^4.
    \label{eq:neutrinoenergydensity2}
\end{equation}
Fig.\ \ref{fig:Neff} shows \neff{} as a function of sterile neutrino lifetime. 

We find that there are two kinds of models that are consistent with CMB-derived observations of \neff. Shorter-lived sterile neutrinos decay early enough such that the active neutrino decay products have ample time to thermalize before decoupling from the plasma, so dilution and the high-energy population of active neutrinos don't significantly affect the final distributions and thus \neff{} is not affected much by these models. Models with steriles that have intermediate lifetimes produce a value of \neff{} that is too low as there is significant dilution of the active neutrinos. Interestingly, in longer lived models, the competing effects of dilution and the production of a population of high-energy, out-of-equilibrium, active neutrinos can balance to result in an \neff{} value consistent with observations. In much longer lifetimes, however, \neff{}  gets very large as far too many high-energy active neutrinos are created. We will compare the two models circled in Fig.\ \ref{fig:Neff}, a long-lived model and a short-lived model. The long-lived models results are more pronounced, so we will explore the results of the long-lived model first so that the nuances of the short-lived model's results are more straightforward.

\subsection{\label{sec:level4a}Long-Lived Sterile Neutrino Decay}

\begin{figure}
    \includegraphics[scale=0.31]{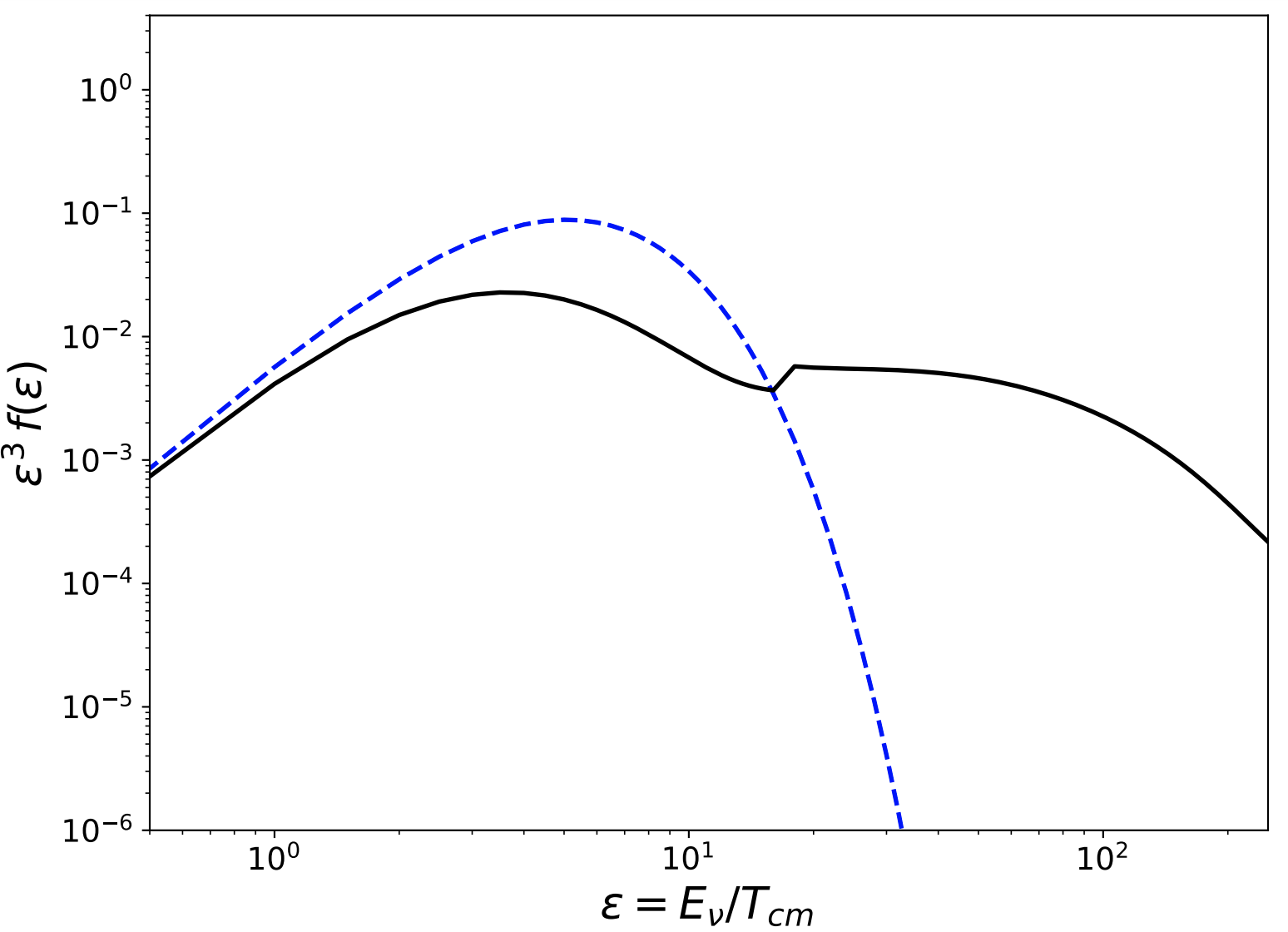}
    \caption{The final spectrum of active neutrinos in the long-lived model, in solid black, is compared to the standard cosmological model in dashed blue. There are significantly fewer low energy active neutrinos due to dilution and there is a population of non-thermal high energy active neutrinos from late time sterile neutrino decays.}
    \label{fig:LongLivedSpectrum}
\end{figure}

Fig. \ref{fig:LongLivedSpectrum} shows the resultant spectrum of active neutrinos and antineutrinos for a sterile neutrino with a lifetime of $\tau=0.91$ s. For comparison purposes, we also included the active neutrino spectrum in the standard cosmological model. In the standard cosmological model, the active neutrino spectrum is thermal, as the neutrinos are in thermal equilibrium with the plasma until they decouple, and thus they maintain the signature thermal spectrum even after decoupling. In our model, however, the spectrum differs from a thermal spectrum in several ways. The most prominent feature is the population of high energy neutrinos. This is attributable to the high-energy active neutrino decay products of the sterile neutrinos. In addition, the effects of dilution can be seen on the low energy side of the spectrum, which presents itself as a lower effective temperature. 

\begin{figure}
    \includegraphics[scale=0.31]{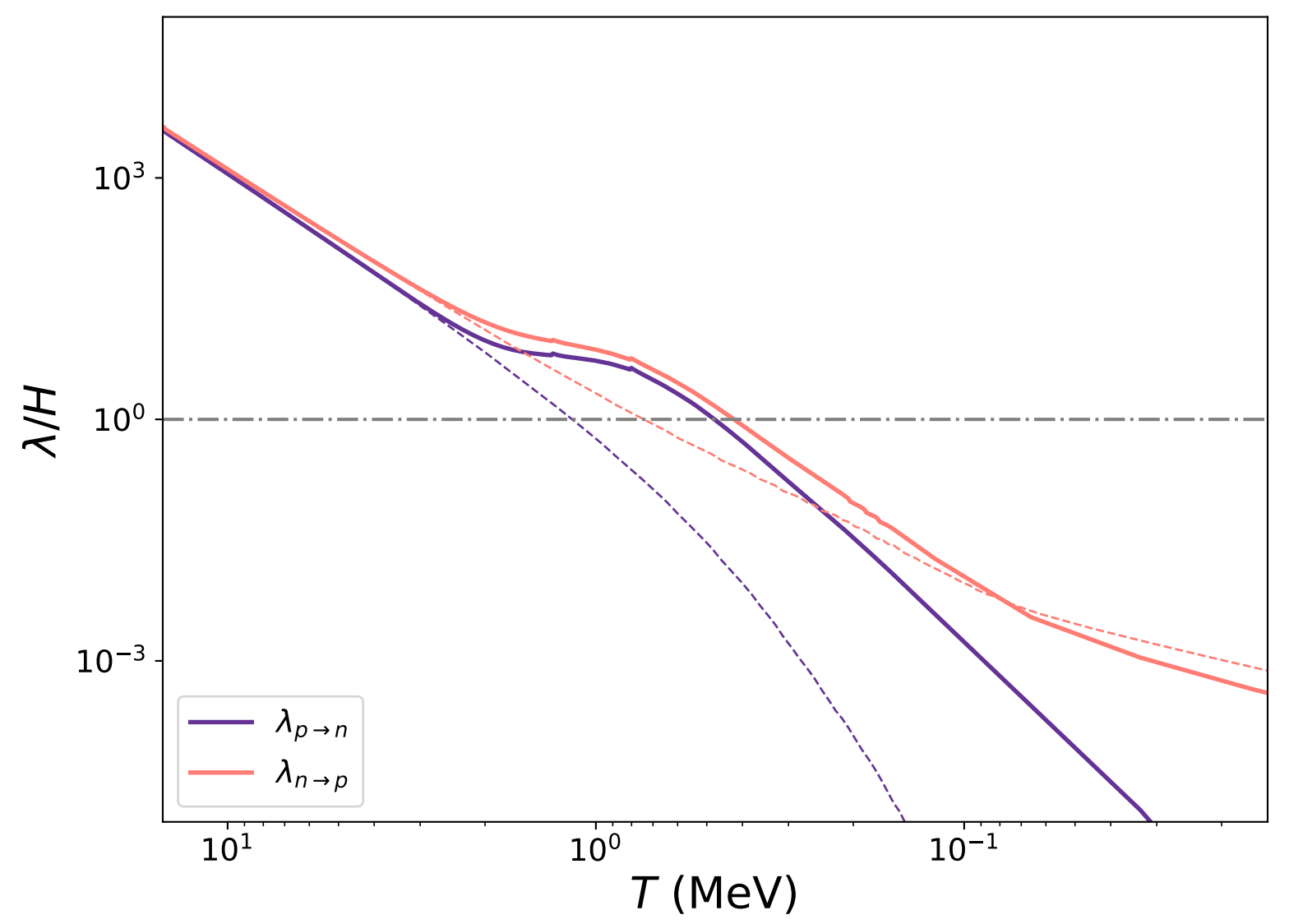}
    \caption{A comparison of the weak neutron/proton interconversion rates through lepton capture in the long-lived model (solid lines) to the standard cosmological model (dashed lines). Specifically, the neutron-to-proton rates (light pink) and proton-to-neutron rates (dark purple) are compared between the two models. The rates are presented as a ratio of these weak rates to the Hubble rate, so the dashed line where $\lambda=H$ serves as a rough guide to the eye to where these processes freeze out.}
    \label{fig:N2PLongLived}
\end{figure}

Fig.\ \ref{fig:N2PLongLived} shows the significant effect that the altered neutrino and antineutrino spectra have on the lepton capture rates \cite{GF16}. Both rates in our model, shown in solid lines, are significantly higher than in the Standard Model, shown in dotted lines. We see that the onset of weak freeze out (roughly where $\lambda_{n \to p}, \lambda_{p \to n} \sim H$) is delayed which will significantly affect the evolution of the neutron-to-proton ratio. These rates are boosted by the population of high energy neutrinos because of the energy dependence of the weak interaction and because the high energy neutrinos easily overcome the threshold for $\bar{\nu}_e$ capture on protons.

\subsection{\label{sec:level4b}Short-Lived Sterile Neutrino Decay}

\begin{figure}
    \includegraphics[scale=0.30]{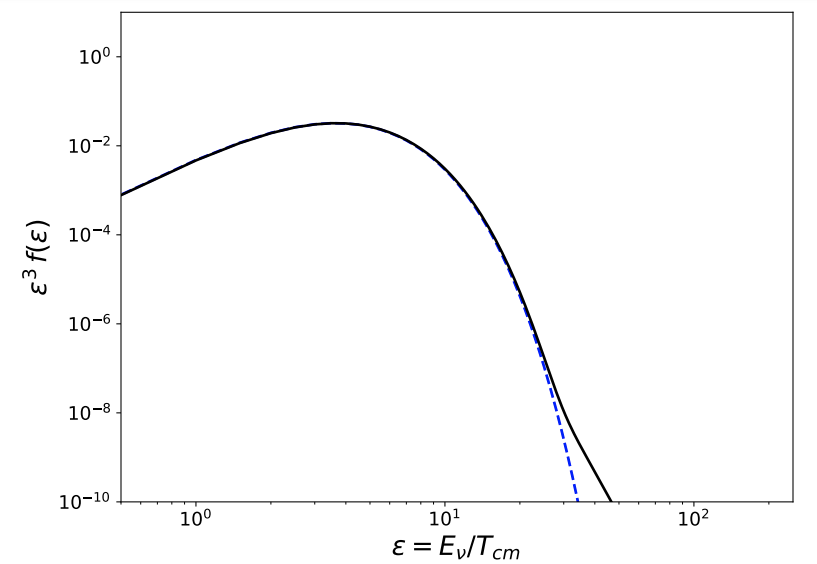}
    \caption{The final spectrum of active neutrinos in the short-lived model ($m_s=300$ MeV, $\tau_s = 0.03$ s, see Sec.\ \ref{sec:level4b}), in solid black, is compared to the standard cosmological model in dashed blue. There is a slight high energy tail in this spectrum from sterile neutrino decays.}
    \label{fig:ShortLivedSpectrum}
\end{figure}

Fig.\ \ref{fig:ShortLivedSpectrum} shows the resultant spectrum of active neutrinos and antineutrinos for a sterile neutrino with a lifetime of $\tau=0.03$ s, and looks significantly different from Fig.\ \ref{fig:LongLivedSpectrum}. The same dilution effects on the low-energy neutrinos and the same high-energy decay effects on the high-energy neutrinos are present, but significantly muted, such that the spectrum looks nearly thermal as it would in the standard cosmological model. Such a drastic difference from the long-lived model's spectrum is due to two effects of the shorter lifetime. First, since the steriles decay earlier, the neutrino and antineutrino products have more time to thermalize. Second, the steriles produce a constant spectrum of neutrinos regardless of lifetime, so these active neutrino decay products are less energetic relative to the plasma if the sterile decays earlier. Therefore, the short-lived sterile neutrino has small effects on the active neutrino spectrum at these lower epsilon values.

\begin{figure}
    \includegraphics[scale=0.31]{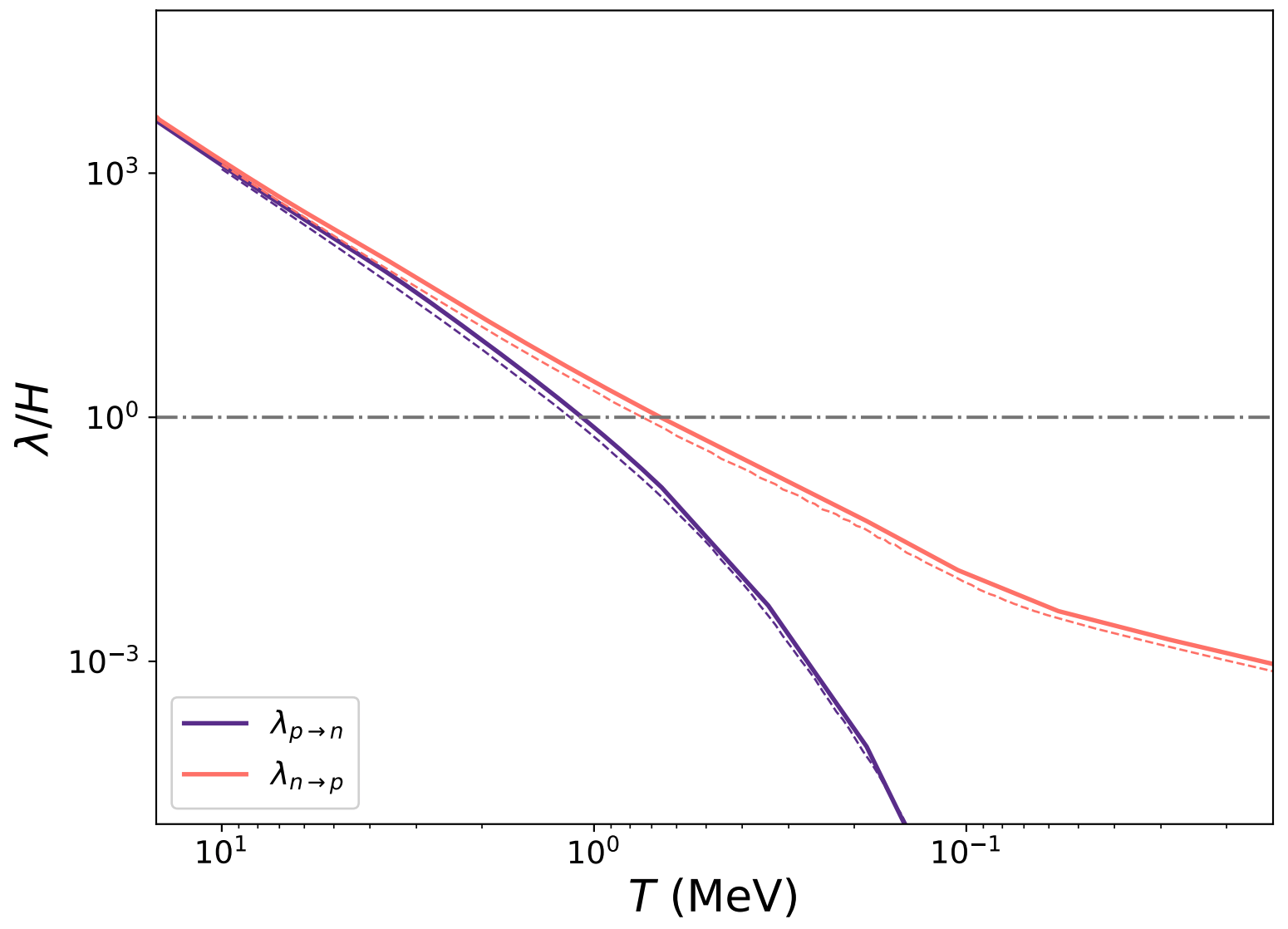}
    \caption{The same comparison of the weak neutron/proton interconversion rates through lepton capture as in Fig.\ \ref{fig:N2PLongLived}, but for the short-lived model. The rates in our model are always at most a few percent above the standard cosmological model rates.}
    \label{fig:N2PShortLived}
\end{figure}

Fig.\ \ref{fig:N2PShortLived} shows the small effect that these perturbations in the active neutrino spectrum have on the lepton capture rates. At most, these represent percent-level increases from their standard cosmological values. These slight differences in rates may affect BBN yields in interesting ways. We look forward to exploring the effects of these models and others like it in future work.

\section{\label{sec:level5}Discussion}

Improved precision in CMB observations can play a useful role in constraining BSM physics. If these observations reveal tension between the observationally-inferred value of \neff{} and the theoretically expected value of 3.046, BSM models such as the one presented here may provide interesting insight into fundamental physics. Even if there aren't discrepancies in these \neff{} values, our models may provide interesting results in the form of effects on BBN yields and the CMB-derived sum of the neutrino masses, $\sum m_{\nu}$.

BBN yields will be affected primarily through the effects of the sterile neutrino decay on the time-temperature relationship and the weak rates that interconvert neutrons and protons. The time-temperature relation is affected by both the presence of an energy density of sterile neutrinos as well as the reheating of the plasma as the sterile neutrinos decay. Furthermore, a high energy population of $\nu_e$ and $\bar{\nu}_e$ will significantly boost the weak rates that govern the neutron-to-proton ratio. In addition, increasing the lifetime of the sterile neutrinos will destroy alpha particles as high-energy decay products can dissociate the alpha particles after they form. In future work, we look to couple our model to a BBN nucleus reaction network to explore the consequences of these effects on BBN yields. 

In addition, our model would impact the interpretation of the cosmological inference of the sum of the neutrino masses, $\sum m_{\nu}$. While the current upper bound on the CMB-inferred value of $\sum m_{\nu}$ is consistent with neutrino oscillation results for the normal and inverted hierarchies, improved sensitivity may either produce an observed signal or upper bound inconsistent with the inverted hierarchy ($\sum m_{\nu} \lesssim $ 105 meV) or perhaps even the normal hierarchy ($\sum m_{\nu} \lesssim $ 58 meV) \cite{CMBS4}. Furthmore, as laboratory-based neutrino mass measurements become more precise, one could imagine a scenario where the CMB and laboratory-based measurements of neutrino mass are inconsistent with each other. In such a scenario, a likely culprit would be the assumptions of the standard cosmological model implicit in the cosmologically-inferred value, especially the assumption that the relic neutrinos have thermal spectra with a temperature that is $(4/11)^{1/3}$ the plasma temperature. The slight high-energy tail in the neutrino distribution of the short-lived model would have a negligible impact on the CMB-derived value of $\sum m_{\nu}$. However, the significantly non-thermal spectrum in the long-lived model would likely significantly reduce the CMB-inferred value of $\sum m_{\nu}$. This would occur due to both a smaller contribution by neutrino mass to the matter density at the current epoch and longer neutrino free-streaming lengths. An interesting hallmark of the long-lived model would be an \neff{} value consistent with expectations of the standard cosmological model but a $\sum m_{\nu}$ that is too small to be consistent with laboratory results.

\acknowledgements
HR, AM, and CK acknowledge support from NSF grant PHY-1812383 and from the College of Arts and Sciences at the University of San Diego.  GMF acknowledges NSF Grant No. PHY-1914242 at UCSD and the NSF N3AS Physics Frontier Center, NSF Grant No. PHY-2020275, and the Heising-Simons Foundation (2017-228). 

\appendix\section{Neutrino Product Decay Spectra}

To incorporate the spectrum of neutrinos and antineutrinos from the decay of sterile neutrinos, we need to determine $(df/da)_{\nu_s\rm\,decay}$. The number of sterile neutrinos in a comoving volume is $n_sa^3$. If the rate that a sterile neutrino decays into an active neutrino with energy between $E$ and $E+dE$ is $dP/dt$, then
\begin{equation}
    \frac{dP}{dt}n_sa^3 = \frac{d}{dt}\left( f(E) \frac{E^2dE}{2\pi^2}a^3\right)
    \label{eq:dfda2}
\end{equation}
is the rate at which such neutrinos are produced through sterile neutrino decay. The RHS of Eq. (\ref{eq:dfda2}) is the time rate of change of the number of neutrinos with energy between $E$ and $E+dE$ in a comoving volume. This can be manipulated to show that
\begin{equation}
    \left. \frac{df}{da} \right\vert_{\nu_s\rm\,decay} = \frac{2\pi^2}{\epsilon^2 T_{\rm cm}^2}\frac{dP}{dtdE}n_s(t)\frac{dt}{da}.
    \label{eq:dfda3}
\end{equation} 
A couple of things should be noted in this result. The distribution function is written as a function of scaled energy, $\epsilon = E/T_{\rm cm}$, which is convenient because $f(\epsilon)$ is not affected by the expansion of the Universe, while $f(E)$ is affected as the Universe expands and cools. The quantity $dP/(dt \, dE)$ is the rate of sterile neutrino decay producing a neutrino with energy between $E$ and $E+dE$, and is written as a function of energy because it only depends on the decay process and sterile neutrino mass. In this Appendix, we introduce $dP/(dt \, dE)$ for the decay processes of interest in his work.

We note that in this work we do not distinguish between $\nu_e$, $\bar{\nu}_e$, $\nu_{\mu}$, and $\bar{\nu}_{\mu}$ in the decays or scattering. While we expect that this simplification affects the specifics of the model described here, we feel the general properties of the results are not greatly affected. We leave this more computationally expensive exercise to future work, especially as we explore BBN in depth where the differences become important.

\subsection{\label{sec:level6a}Neutrino Type I $\nu_s \to \nu + x$}
Two-body decay processes produce mono-energetic active neutrino decay products. The neutrino energy from this two-body decay can be determined kinematically in the sterile neutrino rest frame as
\begin{equation}
    E_\nu = \frac{{m_s}^2 - {m_x}^2}{2m_s},
    \label{eq:EB}
\end{equation}
with $m_{\nu}=0$. Both decay 1 and decay 2 follow this kind of decay process, where product $x$ is a photon in decay 1 and a neutral pion in decay 2. Since the products of decay 1 are both massless, 
\begin{equation}
    E_{\nu1} = \frac{m_s}{2},
    \label{eq:EB1}
\end{equation}
and the energy of the neutrino from decay 2 is 
\begin{equation}
    E_{\nu2} = \frac{m_s^2-m_{\pi^0}^2}{2m_s}.
    \label{eq:EB2}
\end{equation}
Therefore, active neutrino products of these decays will exclusively produce neutrinos with a given energy, so it follows that $dP/(dt\,dE) \propto \delta(E-E_I)$ where $E_I$ is $E_{\nu1}$ for decay 1 or $E_{\nu2}$ for decay 2. $\Gamma_0$ is the decay rate of the sterile neutrino decay of interest, and can be written as
\begin{equation}
    \Gamma_0 = \int^{\infty}_0 \frac{dP}{dtdE}dE.
    \label{eq:Gamma0}
\end{equation}
Therefore, we can say that for these type I decays, 
\begin{equation}
    \left(\frac{dP}{dtdE}\right)_{I} = \Gamma_{0} \, \delta\left(E-E_I\right).
    \label{eq:dPdtdEI}
\end{equation}

\subsection{\label{sec:level6b}Neutrino Type II $\nu_s \to \pi^{\pm} \to \nu_\mu + \mu^\pm$}
The active neutrinos produced from decays 3 and 4, however, are not so easily described. Let's begin with the active neutrino product from the charged pion decay of these two processes. The pion is monoenergetic in the rest frame of the sterile neutrino, and the active neutrino product of the pion decay is monenergetic in the rest frame of the pion. However, we are interested in the energy of the active neutrino in the rest frame of the sterile neutrino, so we Lorentz transform the energy of the active neutrino back to the rest frame of the sterile neutrino, yielding a range of potential energies of the active neutrino of equal likelihood due to the different directions at which the pion could have decayed into the muon and the active neutrino. For this decay chain, $dP/(dt \, dE)$ is defined by the ``top hat'' function,
\begin{equation}
    \left(\frac{dP}{dtdE}\right)_{II} = \Gamma_{0} \times
    \begin{cases}
        \displaystyle \frac{1}{2\gamma_\pi v_\pi E_\nu '} & \text{if $E_{\rm min}\leq E\leq E_{\rm max}$} \\
        0 & \text{otherwise} 
    \end{cases} .
    \label{eq:dPdtdEII}
\end{equation}
The minimum and maximum energies of the active neutrino are $E_{\rm min,max}=\gamma_\pi(E_{\nu}' \pm v_\pi p_{\nu}')$, the Lorentz factor between the pion's rest frame and the rest frame of the sterile neutrino is $\gamma_\pi = E_\pi/m_\pi$, and the speed of the pion in the sterile neutrino rest frame is $v_\pi=p_{\pi}/E_{\pi}$. The energy and momentum of the active neutrino in the rest frame of the pion, $E_{\nu}'$ and $p_{\nu}'$, can be derived from the 2-body decay,
\begin{equation}
    E_{\nu}' = \frac{m_{\pi^{\pm}}^2 - m_{\mu}^2}{2m_{\pi^{\pm}}}.
\end{equation}

\subsection{\label{sec:level6c}Neutrino Type III $ \nu_s \to \mu^\pm \rightarrow e^\pm + \nu_e + \bar{\nu}_\mu$ }
In decay 4, the sterile neutrinos decay into a charged pion and a muon, and the active neutrinos produced by the decay of this muon must be considered as well. The differential decay rate measured in the rest frame of the muon, $d\Gamma_{\mu}'/dE_{\nu}'$, is a well-known spectrum with neutrino energies between $0$ and $E_{\rm max}' = m_\mu(1-(m_e/m_\mu)^2)/2$ \cite{Fetscher1992,Greub1994}. This neutrino spectrum is formed from a distribution of monoenergetic muons, as measured in the sterile neutrino rest frame, so to Lorentz transform this distribution into the sterile neutrino rest frame, we use the results of the previous section (the ``top hat'' distribution). The rate,
\begin{equation}
\begin{split}
    \Gamma_{\mu_a}' &= \int_a^{b} \frac{d\Gamma_\mu'}{dE_\nu'} \frac{dE_\nu'}{E_\nu'} \\
    &= \frac{8G_Fm_\mu^2}{16\pi^3} \int_a^b \frac{E_\nu'}{1-\frac{2E_\nu'}{m_\mu}}\left(1-\frac{2E_\nu'}{m_\mu}-\frac{m_e^2}{m_\mu^2}\right)^2 dE_\nu'.
    \end{split}
    \label{eq:Gammamua}
\end{equation}
is the rate, measured in the muon rest frame, that muons decay into neutrinos with energy $E$, as measured in the sterile neutrino rest frame. The limits of integration are $a=E/(\gamma_\mu(1+v_\mu))$ and $b={\rm min}(E_{\rm max}',E/(\gamma_\mu(1-v_\mu)))$ and $\gamma_\mu = E_\mu/m_\mu$ and $v_{\mu} = p_{\mu}/E_{\mu}$ are the Lorentz factor and speed of the muon's rest frame relative to the sterile neutrino rest frame.

However, muon decay is fast compared to the sterile neutrino decay we explore here, so in the production of active neutrinos, the rate-limiting step is the sterile neutrino decay. So, normalizing for the muon decay rate (in its rest frame),
\begin{equation}
    \left(\frac{dP}{dtdE}\right)_{III} = \frac{\Gamma_{0}\Gamma_
    {\mu_a}'}{2\gamma_{\mu}v_{\mu}\Gamma_{\mu_b}'}
    \label{eq:dPdtdEIII}
\end{equation}
where the muon decay rate measured in its rest frame is 
\begin{equation}
\begin{split}
    \Gamma_{\mu_b}' &= \int_0^{E_{\rm max}'} \frac{d\Gamma_\mu'}{dE_\nu'} dE_\nu' \\
    &= \frac{8G_Fm_\mu^2}{16\pi^3}\int_0^{E_{\rm max}'} \frac{{E_\nu'}^2}{1-\frac{2E_\nu'}{m_\mu}}\left(1-\frac{2E_\nu'}{m_\mu}-\frac{m_e^2}{m_\mu^2}\right)^2 dE_\nu'.
    \end{split}
    \label{eq:Gammamub}
\end{equation}

\subsection{\label{sec:level6d}Neutrino Type IV $\nu_s \to \pi^{\pm} \to \mu^\pm \rightarrow e^\pm + \nu_e + \bar{\nu}_\mu$}

In both decays 3 and 4, an additional two neutrinos are produced from the decay of the muon that was produced by the pion decay. The crucial difference between this decay and decay type III is that in decay type III the decaying muon in question is monoenergetic in the frame of the sterile neutrino, while the decaying muon in decay type IV is monoenergetic in the frame of the pion, which corresponds to a top hat function distribution for the muon in the sterile neutrino rest frame (see decay Type II). Integrating this spectrum of muon energies over the results from decay type III results in 
\begin{equation}
    \left(\frac{dP}{dtdE}\right)_{IV} = \frac{\Gamma_{0}}{2\gamma_\pi v_\pi p_{\mu}'} \frac{1}{\Gamma_{\mu_b}'} \times \int^{E_{\rm max}}_{E_{\rm min}} \frac{\Gamma_
    {\mu_a}'}{2\gamma_{\mu}v_{\mu}} dE_\mu
    \label{eq:dPdtdEIV}
\end{equation}
where $\Gamma_{\mu_a}'$ and $\Gamma_{\mu_b}'$ are described by Eqs. (\ref{eq:Gammamua}) and (\ref{eq:Gammamub}), respectively. The minimum and maximum energies of the muon in the pion's rest frame are $E_{\rm min,max}=\gamma_\pi(E_{\mu}' \pm v_\pi p_{\mu}')$ where $E_{\mu}'$ and $p_{\mu}'$ are the energy and momentum of the muon in the rest frame of the pion.

%\nocite{*}
\bibliography{refs}

\end{document}